\documentclass[twocolumn]{aa}
\usepackage{graphicx}
\usepackage{txfonts}
\usepackage{booktabs}
\usepackage{natbib}
\usepackage[utf8]{inputenc}
\usepackage{multirow}
\bibpunct{(}{)}{;}{a}{}{,} 

\usepackage[usenames,dvipsnames]{color}
\usepackage{colortbl,xcolor}

\def\sqiggt{\hbox{\rlap{\lower.55ex \hbox {$\sim$}}\kern-.05em \raise.4ex \hbox{$>$}\,}}
\def\sqiglt{\hbox{\rlap{\lower.55ex \hbox {$\sim$}}\kern-.05em \raise.4ex \hbox{$<$}\,}}

\begin{document}

\title{The long-lasting optical afterglow plateau of short burst GRB 130912A}

\author{Biao Zhu$^{1,2}$, Fu-Wen Zhang$^{1,3}$, Shuai Zhang$^{2,4}$, Zhi-Ping Jin$^{2}$, and Da-Ming Wei$^{2}$
}

\institute{College of Science, Guilin University of Technology, Guilin 541004, China\\
\email{fwzhang@hotmail.com (FWZ) and jin@pmo.ac.cn (ZPJ)}
\and
Key Laboratory of Dark Matter and Space Astronomy, Purple Mountain Observatory, Chinese Academy of Sciences,
Nanjing, 210008, China.
\and
Key Laboratory for the Structure and Evolution of Celestial
Objects, Chinese Academy of Sciences, Kunming 650011, China
\and
Graduate University of Chinese Academy of Sciences, Beijing 100049, China.
}

\date{Received XXXX; accepted XXXX}

\authorrunning{Zhu et al.}

\titlerunning{optical plateau of GRB 130912A}

\abstract
{The short burst GRB 130912A was detected by {\it Swift}, Fermi satellites, and several ground-based optical telescopes. Its X-ray light curve decayed with time normally. The optical emission, however, displayed a long-term plateau.
}
{We examine the physical origin of the X-ray and optical emission of short GRB 130912A.}
{The afterglow emission was analysed and the light curve fitted numerically.}
{The canonical forward-shock model of the afterglow emission accounts for the X-ray and optical data self-consistently, so the energy injection model that has been widely adopted to interpret the shallowly decaying afterglow emission is not needed.}
{The burst was born in a very-low density interstellar medium, which is consistent with the compact-object merger model. Significant amounts of the energy of the forward shock were given to accelerate the non-thermal electrons and amplify the magnetic fields (i.e., $\epsilon_{\rm e}\sim 0.37$ and $\epsilon_{\rm B}\sim 0.16$, respectively), which are much more than those inferred in most short-burst afterglow modelling and can explain why the long-lasting optical afterglow plateau is rare in short GRBs.}


\maketitle

\section{Introduction}
Gamma-ray bursts (GRBs) are short flashes of $\gamma$-rays from the outer/deep space. Based on the duration distribution of the prompt emission, GRBs can be divided into two sub-groups \citep{1993ApJ...413L.101K}. One group has a typical duration of $\sim 20~{\rm s}$. The other has a much shorter duration $<2$ s, centered at $\sim 0.1$ s. The long GRBs are usually related to the death of massive stars, and one smoking-gun signature of such a scenario is the bright supernova emission, as identified in most nearby long GRBs \citep{1993ApJ...405..273W,1999ApJ...524..262M,2003Natur.423..847H}. The short GRBs are rarer by a factor of about four  according to the BATSE, or about ten from the {\it Swift} observation. The rate depends strongly on the energy bands and on sensitivities of the instruments (e.g., Qin et al. 2013, Zhang et al. 2012). The understanding of such a group of events had not been revolutionized until 2005 when {\it Swift} and {\it HETE-II} localized such events, and the long-lasting multi-wavelength afterglow emission had been detected \citep{2005Natur.437..851G,2005Natur.437..845F}.

Although not as abundant as that of long GRBs, the afterglow emission data of short GRBs by then are valuable for revealing the physical processes taking place in the central engine. For example, the peculiar X-ray emission, such as the X-ray flares and the X-ray plateau followed by abrupt quick decline \citep{2005Natur.437..855V,2005Natur.438..994B,2010MNRAS.409..531R,2011MNRAS.417.2144M}, was found to be inconsistent with the so-called standard forward-shock afterglow model. This emission instead implies the prolonged activity of the central engines \citep{2005ApJ...635L.129F,2006Sci...311.1127D,2006ApJ...639..363P,2006ChJAA...6..513G,2006MNRAS.370L..61P,2008MNRAS.385.1455M,
2013MNRAS.430.1061R}, which is possibly associated with non-ignorable gravitational wave radiation \citep{Fan2013PRD,2013ApJ...763L..22Z}. If these X-ray plateaus are indeed powered the internal energy dissipation of supra-massive neutron star (SMNS) wind and the sharp declines mark the collapse of the SMNSs  \citep{2006ChJAA...6..513G,2013MNRAS.430.1061R}, then the maximum gravitational mass of non-rotating neutron star can be estimated to be $\sim 2.3~M_{\odot}$ \citep{Fan2013PRD,LiX2014}. Moreover, the detection of a weak infrared bump in GRB 130603B strongly favors the physical origin of the merger of a compact object binary \citep{2013Natur.500..547T,2013ApJ...774L..23B,2013ApJ...775L..19J}. To account for the shallowly decaying X-ray emission, a millisecond magnetar central engine, which was still active in $\sim 10^{3}$ s after the short burst, is needed, and the progenitor stars are likely to be double neutron stars \citep{2013ApJ...779L..25F,2014A&A...563A..62D,2014ApJ...785...74L,2015arXiv150102589L}.

Motivated by this progress, in this work we examine the physical origin of the X-ray and optical emission of short GRB 130912A, which is characterized by a long-lasting optical plateau. In section 2 we introduce the observations of GRB 130912A. In section 3 we interpret the data. We summarize our results with some discussion in section 4.

\section{Observations}

\subsection{Swift observations}
The {\it Swift} Burst Alert Telescope (BAT) detected GRB 130912A at 08:34:57 UT on September 12, 2013 \citep{2013GCN..15212...1D}.
The $T_{90}$ duration is $0.28\pm0.03$ second. The light curve  shows two overlapping peaks.
The time-averaged spectrum of the first 0.32 second is best fitted by a simple power law, with photon index  $\Gamma_{\gamma} = 1.20 \pm 0.20$.
The total fluence is $1.7 \pm 0.2 \times 10^{-7}$ erg cm $^{-2}$, and the peak photon flux is $2.2 \pm 0.3$ ph cm$^{-2}$ s$^{-1}$ (Krimm et al. 2013).
All values are in the 15 -- 150 keV energy band.
There is no evidence of extended emission detected in the BAT energy range \citep{2010ApJ...717..411N}, which makes it an unambiguous short GRB.

The Swift X-ray Telescope (XRT) began to observe the field at 08:36:31.7 UT, 93.9 seconds after the BAT trigger. Kennea et al. (2013)
analyzed the initial XRT data and report that the light curve can be modelled with a power-law decay with a decay index of $\alpha=
1.20\pm 0.04$. The spectrum formed from the PC mode data can be fitted with an absorbed power law with a photon
spectral index of $\Gamma_{X} = 1.57^{+0.20}_{-0.16}$ and $N_{\rm H}=1.49^{+0.69}_{-0.25} \times
10^{21}$ cm$^{-2}$, consistent with the Galactic value of $N_{\rm H}=1.2 \times 10^{21}$ cm$^{-2}$ (Kalberla et al. 2005).

The {\it Swift} UVOT took 150 seconds to find the chart exposure 98 seconds after the BAT trigger \citep{2013GCN..15212...1D}.
No optical afterglow within the enhanced XRT position \citep{2013GCN..15217...1B} has been detected in this initial and the subsequence exposures \citep{2013GCN..15229...1C}.

\subsection{Ground-based optical observations}
The field of GRB 130912A was observed at early times by several ground-based optical telescopes, but only two detected the afterglow. GROND started observing at 08:50 UT on September 12, 2013 (16 min after the GRB trigger), and found the afterglow at coordinates RA (J2000) = 03:10:22.23 and Dec. (J2000) = 13:59:48.7 within the Swift XRT error box. Over the first hour-long observation in the $r'$ filter, the afterglow seemed to be constant, with a magnitude (here and throughout this paper, magnitudes are in AB system) of r'=22.0$\pm$0.2 \citep{2013GCN..15214...1T}.
Another telescope, P60, started observing at 8:48 UT on September 12, 2013 (13 min after the Swift trigger) also in $r'$ band, confirmed the GROND observation that the source was almost unchanged in the first hour. It led to the suspicion that the detected source is the host galaxy of GRB 130912A \citep{2013GCN..15222...1C};
however, when RATIR observed the field again started on 2013 September 13.30, the source had faded down to a limited magnitude 23.89 (3-sigma) in SDSS $r$ band, so the previously detected source should be the afterglow \citep{2013GCN..15226...1B}. According to the RATIR observation, the host galaxy is fainter than 23.89, 23.79, 22.86, 22.38, 22.30, and 21.78 magnitudes in $r$, $i$, $Z$, $Y$, $J$, and $H$ bands, respectively \citep{2013GCN..15226...1B}. None of these ground-based observations is publicly available in the form of raw data. We collected the public GCN data as shown in Table 1, and loose limits had been discarded. The magnitudes used in the following analysis were corrected for the Galactic extinction, assuming $E(B-V)=0.28$ \citep{1998ApJ...500..525S} and a ratio of total-to-selective extinction $R_{V}=3.1$, the Galactic extinction in $r'$ band is $A_{r'}=0.78$.

\begin{table*}
\begin{center}
\caption{Optical observations of the field of GRB 130912A} 
\label{table:1} 
\centering 
\begin{tabular}{c c c c c c} 
\hline\hline 
    Telescope & Data start & Observation time after trigger & Filter$^{\rm a}$ & Magnitude$^{\rm b}$ & Flux \\ 
          &     (UT)      &      (s)  &  & & (erg~cm$^{-2}$s$^{-1}$Hz$^{-1}$) \\
\hline 
    GROND & 08:50/12/09/2013 & 960   & GROND $r'$ & 22.0  $\pm$ 0.20$^{(1)}$ & $ 5.75\times10^{-29} $ \\
    P60   & 08:58/12/09/2013 & 1380 & GROND $r'$ & 21.77 $\pm$ 0.20$^{(2)}$ & $ 7.11\times10^{-29} $ \\ 
    P60   & 09:28/12/09/2013 & 3180  & GROND $r'$ & 22.09 $\pm$ 0.25$^{(2)}$ & $ 5.29\times10^{-29} $ \\
    RATIR & 07:33/13/09/2013 & $\sim8.1\times10^{4}$ & SDSS $r$  & $>$ 23.89$^{(3)}$           & $<1.01\times10^{-29} $\\
\hline 
\end{tabular}
\end{center}

a. The difference between GROND $r'$ and SDSS $r$ is less than 0.04 mag assuming a power-law spectrum with an index $\beta$ between 1 and -2. Here the flux of the afterglow is expressed as $F_{\nu}\propto t^{\alpha}{\nu}^{\beta}$,

b. The flux is reported with the $1\sigma$ statistical error, and the upper limit is at the confidence level of 3$\sigma$.
References:
$^{(1)}$ Tanga et al. (2013),
$^{(2)}$ Cenko et al. (2013),
$^{(3)}$ Butler et al. (2013).

\end{table*}

\section{Interpreting the optical and X-ray afterglow}

In the {\it Swift-}era, long-lasting plateau-like X-ray emission (i.e., the so-called shallow decay phase) was detected well in a good fraction of GRB afterglows (e.g., Zhang et al. 2006; Nousek et al. 2006). The leading interpretation of the long-lasting plateau-like X-ray emission is the energy injection model, which is valid if the central engine works continually, or alternatively, the bulk Lorentz factor of the outflow material has a wide distribution (e.g., Dai \& Lu 1998; Zhang \& M\'esz\'aros 2001; Fan \& Xu 2006). A general prediction of the energy injection model is that the temporal behaviours of multi-wavelength afterglow emission will be shaped simultaneously (Fan \& Piran 2006). However, the X-ray and optical observations of the GRB afterglow usually do not track each other. The X-ray afterglows usually displayed an early shallow decline, which is often not observed in the optical (e.g., Fan \& Piran 2006; Panaitescu et al. 2006). The physical reason is still not clear for such a puzzle, for example, the widely-adopted energy injection model for the X-ray decline is found to be unable to account for the optical data.

Recently, there is another unusual situation. As shown in Fig.1, the X-ray afterglow emission of GRB 130912A can be fitted by a single power law (Beardmore et al. 2013), as found in most GRB X-ray emission. However, the optical emission is plateau-like on a very long timescale $\sim 3.2\times 10^{3}$ s, which is a very strange behaviour that is rarely observed in the optical afterglow. A long-lasting optical plateau was also detected in GRB 090510 (De Pasquale et al. 2010; Gao et al. 2009). The duration of the optical plateau of GRB 090510, however, is just about half of that of GRB 130912A. The optical plateau of GRB 130912A is thus very likely the {\it longest} one people have detected so far in short GRB afterglows. In reality, the lack of optical variability in the first hour after the trigger of GRB 130912A motivated the idea that these emission were from the host galaxy (Tanga et al. 2013; Cenko et al. 2013). The afterglow emission nature of the optical data had not been established until significant fading was identified about one day after the burst (Butler et al. 2013). The main purpose of this work is to interpret the ``unusual" optical plateau and the X-ray emission self-consistently.

\begin{figure*}[t]
\includegraphics[width=10.0cm,angle=0]{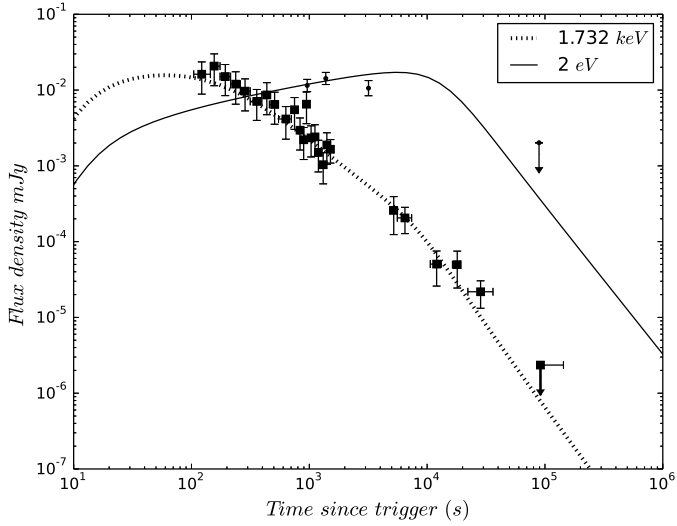}
\caption{X-ray (square) and optical (circle) light curves of GRB 130912A and the theoretical
model curves. The X-ray data are from the UK Swift Science Data Center (Evans et al. 2009) and transformed to 1.732 keV. In order to reduce the influence of the error of the spectral index on the flux calculation, we used the geometric mean of the lower and upper boundaries of the corresponding X-ray energy band $0.3-10$ keV. Proper corrections for extinction in the Milky Way Galaxy have been made. The solid and dashed curves are the theoretical optical and X-ray afterglow prediction with a forward shock.}
\label{fig:Num}
\end{figure*}

Because nothing is unusual in the X-ray band for GRB 130912A, we conclude that the energy injection model does not work for the current data, as discussed above. It is widely known that the forward-shock emission is governed by some physical parameters that can be parameterized as (e.g., Sari et al. 1998; Yost et al. 2003, Fan \& Piran 2006)
\begin{equation}
F_{\nu,{\rm max}}=6.6~{\rm mJy}~\Big({1+z\over 2}\Big) D_{L,28.34}^{-2}
\epsilon_{B,-2}^{1/2}E_{k,53}n_0^{1/2}, \label{eq:F_nu,max}
\end{equation}
\begin{equation}
\nu_m =2.4\times 10^{16}~{\rm Hz}~E_{\rm k,53}^{1/2}\epsilon_{\rm
B,-2}^{1/2}\epsilon_{e,-1}^2 C_p^2 \Big({1+z \over 2}\Big)^{1/2}
t_{d,-3}^{-3/2},\label{eq:nu_m}
\end{equation}
\begin{equation}
\nu_c = 4.4\times 10^{16}~{\rm Hz}~E_{\rm k,
53}^{-1/2}\epsilon_{B,-2}^{-3/2}n_0^{-1}
 \Big({1+z \over 2}\Big)^{-1/2}t_{d,-3}^{-1/2}{1\over (1+Y)^2},
 \label{eq:nu_c}
 \end{equation}
where $C_p \equiv 13(p-2)/[3(p-1)]$, $\epsilon_{\rm e}$ ($\epsilon_{\rm B}$) is the fraction of shock energy given to the electrons (magnetic field), $t_{d}$ is the time in days, the Compton parameter $Y\sim
(-1+\sqrt{1+4\eta \epsilon_e/\epsilon_B})/2$, $\eta \sim \min\{1,
(\nu_m/\bar{\nu}_c)^{(p-2)/2} \}$, and $\bar{\nu}_c=(1+Y)^2 \nu_c$. Here and throughout this text, the convention $Q_{\rm x}=Q/10^{\rm x}$ has been adopted.

At $t\sim 3.2\times 10^{3}~{\rm s}$, if we have $\min\{\nu_{\rm c},\nu_{\rm m}\}=\nu_{\rm c}\gtrsim 5\times 10^{14}~{\rm Hz}$, the temporal behaviour of the optical emission would be $F_{\nu_{\rm opt}}\propto F_{\nu,{\rm max}}\nu_{\rm c}^{-1/3} \propto t^{1/6}$, and the X-ray emission light curve should be $F_{\nu_{\rm x}}\propto  F_{\nu,{\rm max}}\nu_{\rm c}^{1/2} \propto t^{-1/4}$ in the case of $\nu_{\rm c}<\nu_{\rm x}<\nu_{\rm m}$ or, alternatively, $F_{\nu_{\rm x}}\propto  F_{\nu,{\rm max}}\nu_{\rm c}^{1/2}\nu_{\rm m}^{(p-1)/2} \propto t^{-(3p-2)/4}$ in the case of $\nu_{\rm c}<\nu_{\rm m}<\nu_{\rm x}$. While the temporal behaviour of optical emission agrees nicely with the data, the X-ray emission does not. The case of $\nu_{\rm c}<\nu_{\rm x}<\nu_{\rm m}$ is clearly at odds with the data. The case of $\nu_{\rm c}<\nu_{\rm m}<\nu_{\rm x}$ on the timescale of $100~{\rm s}<t<3.2\times 10^{3}~{\rm s}$ is also inconsistent with the X-ray spectrum $F_\nu \propto \nu^{-0.50\pm 0.16}$ (http://www.swift.ac.uk/xrt$_{-}$spectra/).

If we, instead, have $\min\{\nu_{\rm c},\nu_{\rm m}\}=\nu_{\rm m}\gtrsim 5\times 10^{14}~{\rm Hz}$ at $t\sim 3.2\times 10^{3}~{\rm s}$, the temporal behaviour of the optical emission would be $F_{\nu_{\rm opt}}\propto F_{\nu,{\rm max}}\nu_{\rm m}^{-1/3} \propto t^{1/2}$, and the X-ray emission light curve should be $F_{\nu_{\rm x}}\propto  F_{\nu,{\rm max}}\nu_{\rm m}^{(p-1)/2} \propto t^{-3(p-1)/4}$ in the case of $\nu_{\rm m}<\nu_{\rm x}<\nu_{\rm c}$ or, alternatively, $F_{\nu_{\rm x}}\propto  F_{\nu,{\rm max}}\nu_{\rm c}^{1/2}\nu_{\rm m}^{(p-1)/2} \propto t^{-(3p-2)/4}$ in the case of $\nu_{\rm m}<\nu_{\rm c}<\nu_{\rm x}$. Now the temporal behaviours of both optical and X-ray emission are consistent with the data for $p\sim 2.3$, as are the spectral behaviours. Since an optical plateau lasting a few thousand seconds is rare in the short GRB afterglow data, it is highly necessary to examine whether the required forward-shock physical parameters are reasonable or not.

To self-consistently interpret the optical and X-ray data, $\nu_{\rm c}(t\sim 1.0\times 10^{4}~{\rm s})\approx$ $ 7.254\times 10^{16}$ Hz, $\nu_{\rm m}(t\sim 3.2 \times 10^{3}~{\rm s})\approx 5\times 10^{14}$ Hz and $F_{\nu,\rm max}\sim 0.02$ mJy are needed. Following Zhang et al. (2015), we have
\begin{equation}
\epsilon_{\rm B,-2}^{1/2}E_{\rm k,53}n_0^{1/2}\approx a,
\label{eq:4}
\end{equation}
\begin{equation}
E_{\rm k,53}^{1/2}\epsilon_{\rm B,-2}^{1/2}\epsilon_{\rm e,-1}^2 \approx b,
\label{eq:5}
\end{equation}
\begin{equation}
E_{\rm k,53}^{-1/2}\epsilon_{\rm B,-2}^{-3/2}n_0^{-1}{(1+Y)^{-2}}\approx c,
\label{eq:6}
\end{equation}
where
$a=\frac{1}{6.6}F_{\nu,{\rm max}} D_{L,28.34}^{2}\Big({1+z\over 2}\Big)^{-1}$,
$b=\frac{1}{2.4}\times 10^{-16}\nu_m C_p^{-2} \Big({1+z \over 2}\Big)^{-1/2} t_{d,-3}^{3/2}$, and
$c=\frac{1}{4.4}\times 10^{-16}\nu_c\Big({1+z \over 2}\Big)^{1/2}t_{d,-3}^{1/2}$.

Now we have three relations but four free parameters (i.e., $E_{\rm k},~\epsilon_{\rm e},~\epsilon_{\rm B},~n_0$), which implies that these parameters cannot be uniquely determined. However, $(E_{\rm k},~\epsilon_{\rm e},~\epsilon_{\rm B})$ can be expressed as the functions of $n_0$, and it is possible to reasonably constrain the range of $n_0$, as found in Zhang et al. (2015).

In the case of $Y\leq 1$ (i.e., the synchrotron-self Compton cooling is unimportant), $(1+Y)^{-2}$ can be ignored, and we have (see also Zhang et al. 2015)
\begin{equation}
\epsilon_{B,-2}=a^{-\frac{2}{5}}c^{-\frac{4}{5}} n_{0}^{-\frac{3}{5}},
\end{equation}
\begin{equation}
\epsilon_{e,-1}=a^{-\frac{1}{5}}b^{\frac{1}{2}}c^{\frac{1}{10}} n_{0}^{\frac{1}{5}},
\end{equation}
\begin{equation}
E_{k,53}=a^{\frac{6}{5}}c^{\frac{2}{5}} n_{0}^{-\frac{1}{5}}.
\end{equation}
These three variables weakly depend on $n_0$.

In the case of $Y\geq 1$ (i.e., the synchrotron-self Compton radiation is important),  we have $(1+Y)^{-2}\approx Y^{-2}$ and then
\begin{equation}
\epsilon_{B,-2}=(10cd)^{-\frac{8}{5p-9}}a^{-\frac{2(p-1)}{5p-9}}b^{-\frac{4(p-1)}{5p-9}} n_{0}^{-\frac{3p+1}{5p-9}},
\end{equation}
\begin{equation}
\epsilon_{e,-1}=(10cd)^{\frac{1}{5p-9}}a^{-\frac{p-2}{5p-9}}b^{\frac{3p-5}{5p-9}} n_{0}^{\frac{p-1}{5p-9}},
\end{equation}
\begin{equation}
E_{k,53}=(10cd)^{\frac{4}{5p-9}}a^{\frac{6p-10}{5p-9}}b^{-\frac{2(p-1)}{5p-9}} n_{0}^{-\frac{p-5}{5p-9}},
\end{equation}
where $d= \big (\frac{2.4}{4.4} C_{p}^{2}(\frac{1+z}{2})t_{d,-3}^{-1} \big )^{\frac{p-2}{2}}$.
Compared with the case of $Y\leq 1$, $\epsilon_{\rm B}$ and $E_{k}$ depend strongly on $n_0$.

\begin{figure*}[t]
\includegraphics[width=10.0cm,angle=0]{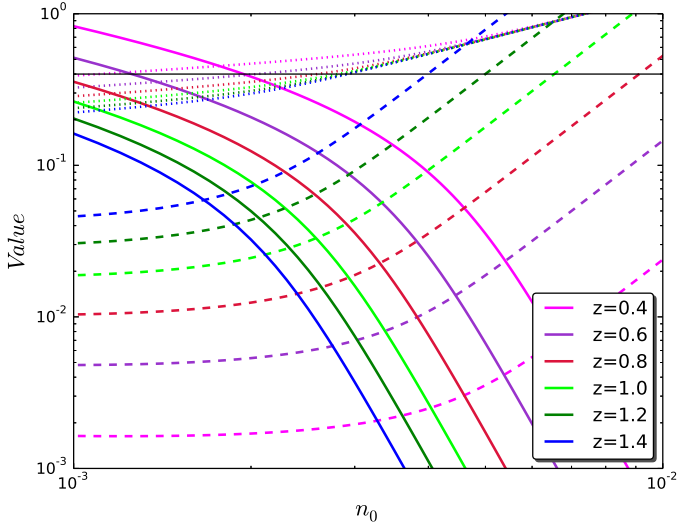}
\caption{$(E_{\rm k},~\epsilon_{\rm e},~\epsilon_{\rm B})$ as functions of $n$ obtained in solving eqs.(\ref{eq:4}-\ref{eq:6}). The solid, dotted, and dashed lines represent $\epsilon_{\rm B}$, $\epsilon_{\rm e}$, and $ E_{\rm k,53}$, respectively. The thin black line (i.e., ${\rm Value}=0.4$) is the {\it reasonable} upper limit of $\epsilon_{B}$ and $\epsilon_{e}$, above which the solution is unphysical.}
\label{fig:Ana}
\end{figure*}

The redshift $z$ is unknown, and we assume that it is in the reasonable range of $0.1\leq z\leq 1.4$. However, for $z<0.4$ in solving eqs.(\ref{eq:4}-\ref{eq:6}), we do not find any reasonable values of the parameters with any given $n_0$, while solutions are obtainable for larger $z$. In Fig. \ref{fig:Ana} we present the physical parameters $(E_{\rm k},~\epsilon_{\rm e},~\epsilon_{\rm B})$ as functions of $n_0$ for $z=(0.4,~0.6,~0.8,~1.0,~1.2,~1.4)$, respectively. As already shown in the analytical approaches, both $E_{\rm k}$ and $\epsilon_{\rm B}$ evolve with $n_0$ quickly, while the dependence of $\epsilon_{\rm e}$ on $n_0$ is rather weak. One can also find from the figure that there is a very interesting constraint that $n_0\leq 3\times 10^{-3}~{\rm cm^{-3}}$ for a {\it reasonable} $\epsilon_{\rm e}\leq 0.4$ ( i.e., expected to be not much larger than the equipartition value $\sim 1/3$), though the afterglow physical parameters cannot be uniquely determined. GRB 130912A was therefore born in a very low-density medium, consistent with the compact-object merger model.

To better show whether the afterglow model can indeed reasonably account for the data or not, in Fig.\ref{fig:Num} we numerically fit the optical and X-ray data of GRB 130912A. The numerical calculation code was developed by Fan \& Piran (2006) and \citet{2006ApJ...642..354Z}. In it, (i) the dynamical evolution of the outflow formulated by \citet{2000ApJ...543...90H} that can describe the dynamical evolution of the outflow
for both the relativistic and non-relativistic phases has been adopted. (ii) The energy distribution of the shock-accelerated electrons is calculated
by solving the continuity equation with the power-law source function $Q\propto \gamma_{\rm e}^{-p}$, normalized by a local injection rate
\citep{2000ApJ...529..151M}. (iii) The cooling of the electrons due to both synchrotron and synchrotron-self Compton has been taken into account.
Assuming $z=0.72$ (following \citet{2013MNRAS.430.1061R} we adopt the average value of redshift of short GRBs), the fit parameters are $(E_{\rm k},~n_0,~\epsilon_{\rm e},~\epsilon_{\rm B},~p,~\theta_{\rm j})=(1.7\times 10^{51}~{\rm erg},~0.002,~0.37,~0.16,~2.3,~0.03)$, where $\theta_{\rm j}$ is the half opening angle of the GRB ejecta. An isotropic fireball is found to be unable to reproduce the data. The inferred $\epsilon_{\rm e}$ and $\epsilon_{\rm B}$ are at the high end of the distribution of the shock parameters of short GRB afterglows (e.g., Soderberg et al. 2006; De Pasquale et al. 2010), which is unexpected since the optical afterglow plateau of GRB 130912A has the longest duration people have ever detected in short events. We would like to point out that the above fit parameters are for $A_{r'}=0.78$, which is very high. If $A_{r'}$ is intrinsically smaller, $F_{\nu_{\rm max}}$ and hence $a$ are lowered accordingly. As shown in eqs.(7-9), or alternatively eqs.(10-12), $\epsilon_{\rm B}$ and $\epsilon_{\rm e}$ would increase, while $E_{\rm k}$ would decrease. The contrary holds in the case of larger $A_{r'}$.

\section{Summary and conclusions}
The most remarkable feature of the short burst GRB 130912A is an optical plateau lasting about $4000$ s, which is the longest one in current short GRB observations, and it is about twice longer than that of GRB 090510. In this work we examined whether any "unusual" information can be extracted from the afterglow data of GRB 130912A. Though the energy injection model has been widely adopted to interpret the shallowly decaying afterglow emission of long and short GRBs (see Zhang et al. 2006; Nousek et al. 2006 and the references therein), it was found to be unable to account for the X-ray and optical data of GRB 130912A self-consistently. Instead the canonical afterglow emission of an ejecta with an opening angle $\theta_{\rm j}\sim 0.03$ can reasonably reproduce the data. The circum-burst medium is found to be ISM-like and has a very low density $\sim 10^{-3}~{\rm cm^{-3}}$, consistent with the model of merger of binary compact objects (either double neutron stars or a neutron star$-$black hole). Significant amounts of the energy of the forward shock were given to accelerate the non-thermal electrons and amplify the magnetic fields (i.e., $\epsilon_{\rm e}\sim 0.37$ and $\epsilon_{\rm B}\sim 0.16$, respectively), which are much more than those inferred in most short burst afterglow modelling and which can explain why the long-lasting optical afterglow plateau is rare in short GRBs.

{\it Acknowledgements.} We thank the anonymous referee for helpful comments and Dr. Y. Z. Fan for stimulating discussion. This work made use of data supplied by the UK Swift Science Data Centre at the University of Leicester, and is supported in part by 973 Program of China under grant 2014CB845800, National Natural Science Foundation of China under grants 11163003, 11273063, 11103084, 11303098, U1331101, and 11361140349, and the Chinese Academy of Sciences via the Strategic Priority Research Programme (Grant No. XDB09000000). F.-W.Z. also acknowledges the support by the Guangxi Natural Science Foundation (No. 2013GXNSFAA019002), the Open Research Programme of Key Laboratory for the Structure and Evolution of Celestial Objects (OP201207) and the project of outstanding young teachers' training in higher education institutions of Guangxi.

\end{document}